# IMPROVING THE FERMILAB BOOSTER NOTCHING EFFICIENCY, BEAM LOSSES AND RADIATION LEVELS[*]

I.L. Rakhno, A.I. Drozhdin, N.V. Mokhov, V.I. Sidorov, I.S. Tropin

*Fermi National Accelerator Laboratory, Batavia, Illinois 60510, U.S.A.*

May 14, 2012

## Abstract

A fast vertical 1.08-m long kicker (notcher) located in the Fermilab Booster Long-05 straight section is currently used to remove 3 out of 84 circulating bunches after injection to generate an abort gap. With the maximum magnetic field of 72.5 Gauss, it removes only 87% of the 3-bunch intensity at 400 MeV, with 75% loss on pole tips of the focusing Booster magnets, 11% on the Long-06 collimators, and 1% in the rest of the ring. We propose to improve the notching efficiency and reduce beam loss in the Booster by using three horizontal kickers in the Long-12 section. STRUCT calculations show that using horizontal notchers, one can remove up to 96% of the 3-bunch intensity at 400-700 MeV, directing 95% of it to a new beam dump at the Long-13 section. This fully decouples notching and collimation. The beam dump absorbs most of the impinging proton energy in its jaws. The latter are encapsulated into an appropriate radiation shielding that reduces impact on the machine components, personnel and environment to the tolerable levels. MARS simulations show that corresponding prompt and residual radiation levels can be reduced ten times compared to the current ones.

# IMPROVING THE FERMILAB BOOSTER NOTCHING EFFICIENCY, BEAM LOSSES AND RADIATION LEVELS*

I.L. Rakhno, A.I. Drozhdin†, N.V. Mokhov, V.I. Sidorov, I.S. Tropin,
Fermilab, Batavia, IL 60510, USA.

## Abstract

A fast vertical 1.08-m long kicker (notcher) located in the Fermilab Booster Long-05 straight section is currently used to remove 3 out of 84 circulating bunches after injection to generate an abort gap. With the maximum magnetic field of 72.5 Gauss, it removes only 87% of the 3-bunch intensity at 400 MeV, with 75% loss on pole tips of the focusing Booster magnets, 11% on the Long-06 collimators, and 1% in the rest of the ring. We propose to improve the notching efficiency and reduce beam loss in the Booster by using three horizontal kickers in the Long-12 section. STRUCT [1] calculations show that using horizontal notchers, one can remove up to 96% of the 3-bunch intensity at 400-700 MeV, directing 95% of it to a new beam dump at the Long-13 section. This fully decouples notching and collimation. The beam dump absorbs most of the impinging proton energy in its jaws. The latter are encapsulated into an appropriate radiation shielding that reduces impact on the machine components, personnel and environment to the tolerable levels. MARS [2] simulations show that corresponding prompt and residual radiation levels can be reduced ten times compared to the current ones.

## NEW SYSTEM STUDIES

A fast vertical kicker (notcher) located in the Booster Long-05 straight section is used to remove 3 out of 84 circulating bunches after injection to produce a gap in the longitudinal distribution, which allows beam abort or extraction from the Booster without loss. The notcher length is 1.08m, and maximum magnetic field is 72.5 Gauss. In the calculations, the distribution of five Booster main bunches is simulated by 121 short one-nsec "fictitious bunches" with a distribution shown in Fig. 1: for removed bunches - by red, circulating - by green, and survival after notching - by blue color. The waveform of existing notcher is shown by red dot line.

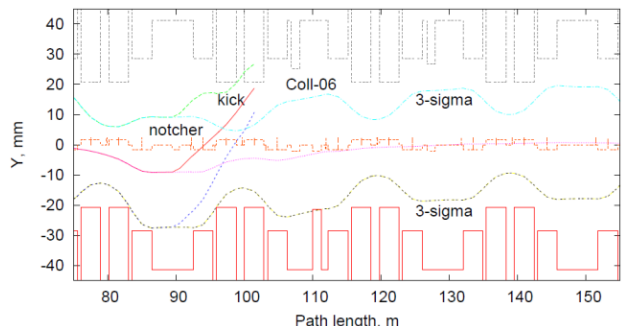

Figure 2: Position of 3-σ circulating and removed beams at notching in a vertical plane.

A vertical position of the 3-σ circulating and removed beams at notching in the Booster-05/06 straight sections are presented in Fig. 2. Circulating beam is at 3σ from the collimator at the Long-06 straight section. According to simulations, 87% of the 3-bunch intensity in total is lost at 400 MeV, with 75% loss on the pole tips of the focusing Booster magnets, 11% on the Long-6 collimator, and 1% in the rest of the ring. The vertical aperture of the Booster Foc and Defoc magnets is very small (21mm and 28mm, respectively) and a 3-sigma beam size is very close to the aperture restriction at injection. This does not permit to dump the beam to the collimator in the vertical plane.

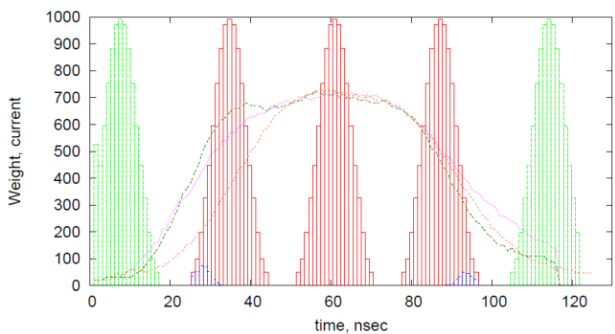

Figure 1: Longitudinal distribution of removing (red), circulating (green) and survival at notching (blue) bunches.

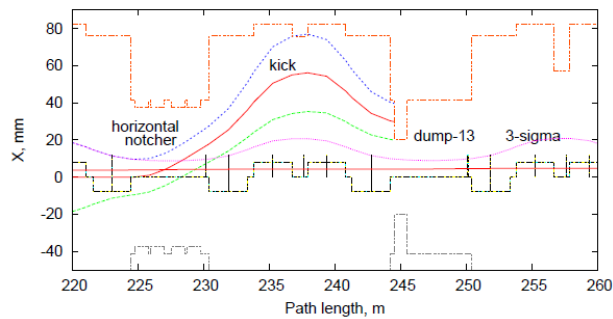

Figure 3: Position of 3-σ circulating and removed beams at notching in a horizontal plane.

The aperture is much larger in a horizontal plane. The position of circulating and removed beams are shown in the Booster-12/13 sections in Fig. 3 with horizontal notching at 700 MeV. Unfortunately, three notchers are required. The fractional parts of the 3 bunch intensity loss for various strengths of 3 notchers is shown in Table 1.

___________________________________________
*Work supported by Fermi Research Alliance, LLC under contract No. DE-AC02-07CH11359 with the U.S. Department of Energy.
† E-mail: drozhdin@fnal.gov

Table 1: Beam loss at horizontal notching.

| Energy,MeV | B,Gauss | Total,% | Dump-13,% | Short-12,% |
|---|---|---|---|---|
| 400 | 62.0 | 98.2 | 92.4 | 6.0 |
| 400 | 55.0 | 95.2 | 94.7 | 0.4 |
| 400 | 45.0 | 84.8 | 84.5 | 0.2 |
| 700 | 85.0 | 97.9 | 95.7 | 2.1 |
| 700 | 80.0 | 96.6 | 96.1 | 0.4 |
| 700 | 72.5 | 93.2 | 93.1 | 0.0 |

The waveform of the existing 1.08-m long notchers at the Long-12 section is shown by red line in Fig. 1. The new design for twice shorter notchers, to be placed instead of the existing ones, shown by blue line, will permit to improve the notcher pulse waveform.

The lost particle population at the beam dump in the Booster Long-13 straight section at 400 and 700 MeV is presented in Figure 4.

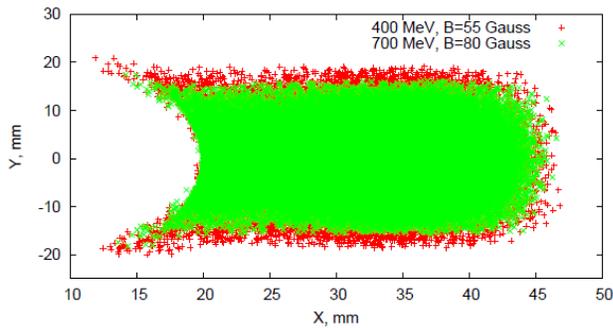

Figure 4: Lost particles population at 400 and 700 MeV in the Long-13 beam Dump.

## PROMPT AND RESIDUAL RADIATION

To investigate consequences of beam losses with the present and proposed notching schemes, the MARS models of the Long-06/13 sections were developed. Geometry of sections were implemented by means of a new MAD/MARS beam builder based on the ROOT [3] libraries. ROOT visualisation of the accelerator elements inside the tunnel are presented on Fig. 5.

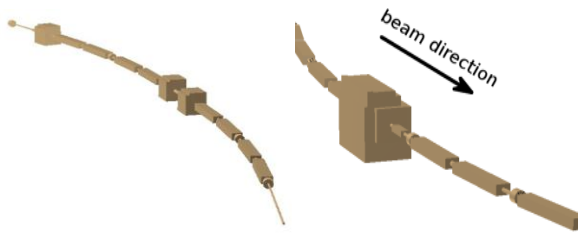

Figure 5: ROOT view of MARS models for the Long-06/07 (left) and 12/13 (right) sections.

Residual dose profiles in the Long-06/07 sections for the present scheme and proposed design are shown in Fig. 6. One can see that residual dose is widely spread in the tunnel with maximum of 1 mSv/h (100 mrem/h) at the surface of magnets. Calculated results for the present scheme are in a good agreement with measurements.

Change in the kick direction from vertical to horizontal for the Long-05 section kickers allows to localize particle fluxes in the region near collimator and decrease activation of the magnets at least by an order of magnitude.

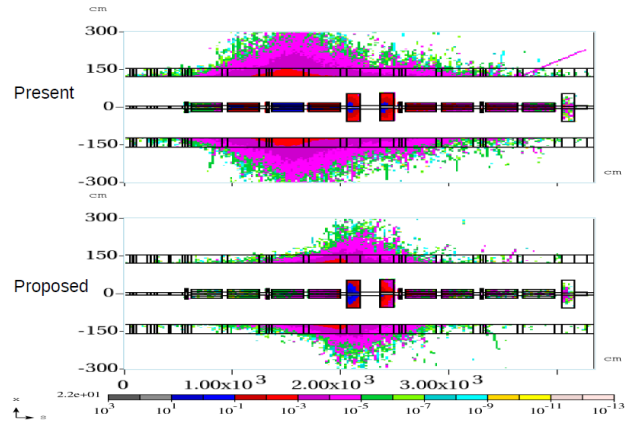

Figure 6: Residual dose profiles (mSv/h) in the Long-06/07 sections for present (top) and proposed (bottom) schemes for 30-day irradiation and 1-day cooling.

It have to be noted that collimators in the Long-06/07 sections are not designed to serve as beam dumps for notching and do not have appropriate shielding. Taking into account this fact, and intending to fully decouple beam collimation and notching, it was decided to use kickers in the Long-12 section with the new beam dump at Long-13 for notching of $1.8 \times 10^{11}$ protons per pulse at 700 MeV with the 15Hz repetition rate.

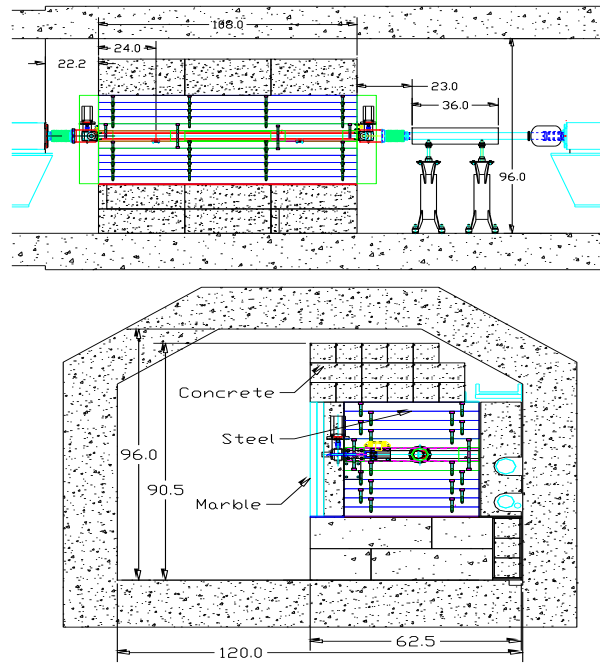

Figure 7: Long-13 section beam dump. Dimensions are in inches.

Energy deposition, residual and prompt doses and ground water activation were calculated and optimized in our analysis by means of the MARS15 code. The final design of the beam dump is shown in Fig. 7, with corresponding results given below.

As shown in Fig. 8, 9 the beam dump allows to keep the maximum contact residual dose on its surface and on surrounding elements below 0.5 mSv/h (50 mrem/h). The peak star density immediately outside of the tunnel wall is safely below of the sump water limit of 4000 cm$^{-3}$ s$^{-1}$.

Analysis of calculated energy deposition shows that temperature, stresses and deformation inside the proposed beam dump assembly will be kept well below the allowable limits.

Prompt dose isocontours in the Long-13 section for the proposed system are shown in Fig.10. To catch high-energy particles escaping the dump jaws and confine radiation in the region, an additional steel mask was implemented immediately downstream the dump. Using these results as a source term, prompt dose level estimated in the Booster gallery and at the dirt surface above Long-13 do not exceed 0.004 mSv/h (0.4 mrem/h).

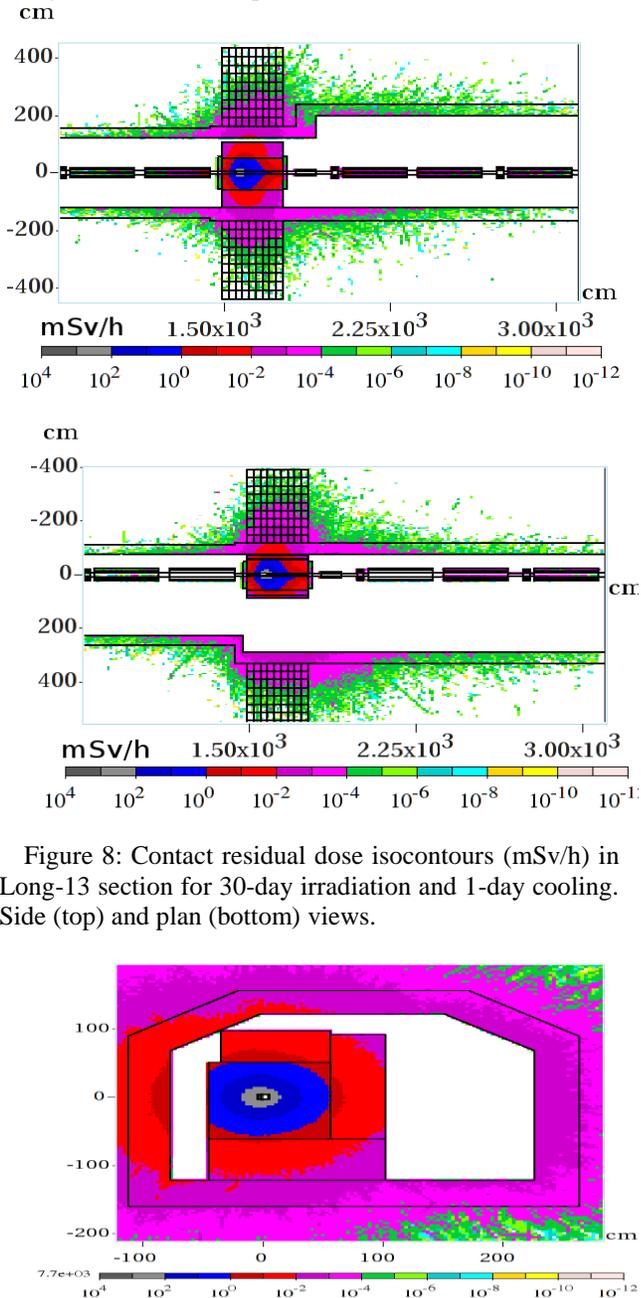

Figure 8: Contact residual dose isocontours (mSv/h) in Long-13 section for 30-day irradiation and 1-day cooling. Side (top) and plan (bottom) views.

Figure 9: Contact residual dose isocontours (mSv/h) at the hottest region of Long-13 section for 30-day irradiation and 1-day cooling.

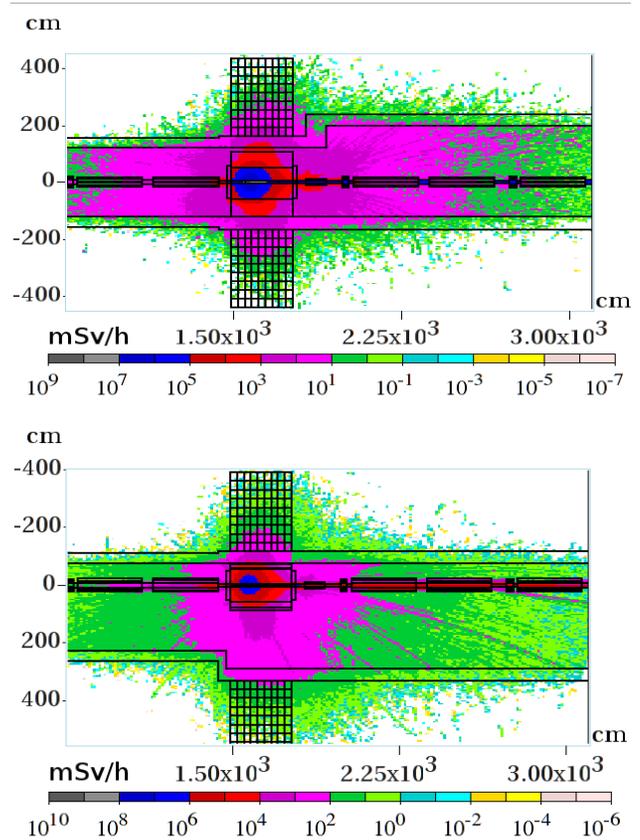

Figure 10: Prompt dose isocontours (mSv/h) in Long-13 section. Side (top) and plan (bottom) views.

## ACKNOWLEDGMENT

The authors acknowledge the helpful discussions and suggestions from W. Pellico and S. Chaurize.